\documentstyle[LTpaper]{article}
\begin{document}
%
% Beginning here, insert your text for the contributed paper
% Title of your contribution:
%
\title{Ground state and excited states of a trapped  dilute condensed Bose gas}
%
% Names of the authors:
%
\author{Alexander L. Fetter}

%
% Institutions and addresses:
\address{\mbox{Departments of Physics and Applied Physics, Stanford
University}\\
\mbox{Stanford, California 94305-4060, USA}}

%
% Abstract of your paper:
%
\abstract{I review recent theoretical treatments of a dilute interacting
condensed Bose gas in a
trap. Bogoliubov's  classic results for a uniform
condensate are generalized to include  the effect of a trap, using the
Gross-Pitaevskii
formalism (for the condensate) and the Bogoliubov equations (for   the
linearized small-amplitude
excitations  of the condensate).  Several recent theoretical studies are
discussed along with some open
questions.\thanks{Research supported in part by the National Science
Foundation under Grant No.~DMR
94-21888.}}
\maketitle\pagestyle{empty}

%
% The main text follows. Put you own text instead of our instructions.
% You may change the names of the sections and subsections in the braces.
%
\section{INTRODUCTION}
Although $^4$He was liquified in 1908 and the anomalous peak in the heat
capacity at $T_\lambda\approx
2.2$ K  was discovered in 1932, its remarkable low-temperature
superfluidity was observed only in 1938
by  Kapitza and by Allen and
Misener \cite{London,Wilks}.  Soon afterward,  Landau's bold 1941
theory of superfluidity of He II (including both the  two-fluid
hydrodynamics and  the
quasiparticle picture of phonons and rotons) \cite{LL,Wilks,FW1}  was
brilliantly confirmed
 by Andronikashvili, Kapitza, and Peshkov \cite{London,Wilks}.
In 1947, Bogoliubov \cite{Bog,Pines,FW2} proposed the first microscopic
description of superfluidity in
a uniform  weakly interacting Bose gas and (with acknowledgment to Landau)
suggested  the correct
generalization to  a dilute gas with strong repulsive interactions.  This
work is summarized in
Sec.~2;  it has since served as the basis for studies of a {\it
nonuniform\/} dilute  Bose gas
\cite{EPG,LPP,FW3}.

The last year has seen
the exciting development of   wholly new Bose condensates consisting of
dilute alkali atoms
\cite{And,Brad,Dav,Mewes}.  At this conference, the F.\ London award has
been presented to Cornell and
Wieman,  with comprehensive invited talks by Wieman, Ketterle, and Jin. The
present work
seeks to build on these experimental achievements  by reviewing the
theoretical situation for a
dilute Bose gas at low temperatures, well below the onset of Bose-Einstein
condensation.  The presence
of the confining (typically harmonic) trap introduces essential new
features that  are
summarized in Secs.~3 and 4  (for the ground state and low-lying excited
states, respectively).
Section 5 considers  various specific calculations that have already been
performed, and Sec.~6
provides a partial list of open interesting questions.

\section{REVIEW OF BOGOLIUBOV THEORY}
Five decades ago, Bogoliubov \cite {Bog,FW2} proposed a simple and
intuitive description of
a uniform dilute Bose gas, focusing particularly on how the  interacting
ground state  differs from
 that of an ideal Bose gas \cite{London}.  Since the repulsive interactions
favor a uniform
configuration, Bogoliubov  used periodic boundary conditions to ensure that
the unperturbed ideal-gas
single-particle ground state  is the uniform  Fourier component with $\bf
k~=~0$.  This choice is
necessary because, for
other boundary conditions,
  the unperturbed ground state would not be even approximately uniform;
in the thermodynamic limit of
$N$ particles in a volume $V$ (keeping the particle density $n \equiv N/V$
fixed), the introduction of
weak repulsive  interactions produces a dramatic change in any nonuniform
equilibrium density profile
(for example, an ideal Bose gas in a box with rigid walls has a density
proportional to products of
squared real trigonometric functions, whereas the interacting density is
essentially uniform except very
near the walls).

An ideal Bose gas at zero temperature has all particles in the condensate,
so that the number
$N_0$ of particles with zero momentum is just the total
$N$; for an interacting system, however,
$N_0$ is smaller than $N$  (in lowest-order perturbation theory, two
particles can scatter out of the condensate and occupy the many
zero-total-momentum states with
separate momenta
$\bf k$ and $-\bf k$). For a {\it
dilute\/} interacting Bose gas with a macroscopic condensate ($N_0\gg 1$),
Bogoliubov made the  very
simple observation that the {\it commutator\/} of the second-quantized
condensate operators
$[c_0,c_0^\dagger] = 1$ is small compared to their separate effect on the
ground state (which yields a
factor of order
$\sqrt{N_0}\gg 1$).  Thus these condensate operators can be treated as
simple numbers.
In addition, Bogoliubov assumed that  most of the particles remain in
the condensate, with
$N'\equiv N-N_0 \ll N$ so that the total  depletion of the condensate
remains small.  This condition
means that it is sufficient to keep only terms of second order in the
noncondensate operators, and a
linear (canonical) transformation can diagonalize the resulting approximate
hamiltonian.

For plane-wave states (\(\propto e^{i{\bf k\cdot r}}\)) at $T = 0$ K, a
particle with wave vector
$\bf k$ has   two characteristic energies: the kinetic energy
\begin{equation}T_k = \hbar^2k^2/2m,\end{equation}
and the mean ``Hartree''    interaction  energy  with the
macroscopic  condensate
\begin{equation}V_H \equiv n_0V_0,\end{equation}
where $V_0$ is the spatial integral of the interparticle potential (namely
the $\bf k = 0$ component
of the Fourier transform);  note that $V_H$ is effectively a constant
optical potential affecting the
propagation through the medium.  As proposed by Bogoliubov/Landau and
proved subsequently by Lee,
Huang, and Yang
\cite{Bog,Pines}, the correct generalization for a strong repulsive
interparticle potential (whose
Fourier transform diverges)  is to replace
$V_0$ by $4\pi a \hbar^2/m$, where $a$ is the $s$-wave scattering length;
in this case,  the Hartree
potential becomes $V_H = 4\pi a \hbar^2n_0/m$.

 One of Bogoliubov's central results is  the energy
eigenvalue  for a plane wave:
\begin{equation}E_k = \sqrt{2V_HT_k + T_k^2}.\label{en}\end{equation}
Here, the first term under the square root is proportional to $k^2$ and the
second to $k^4$, so that
$E_k$ has the following limits
\begin{equation}E_k \approx \cases{\sqrt{4\pi a n_0}\,\hbar^2k/m,&for $k \to
0$;\cr\noalign{\smallskip}
                                   \hbar^2k^2/2m,&for $k \to
\infty$.\cr}\end{equation}
In particular, the long-wavelength dispersion relation is simply that of a
sound wave (a phonon) with
propagation speed $s \equiv \sqrt {4\pi a n_0}\,\hbar/m$, given by the
appropriate thermodynamic
compressibility \cite{Pines}.  This characteristic linear long-wavelength
spectrum  reflects the
presence of  the  Bose condensate   (note that $s^2\propto n_0$) and differs
qualitatively from that of a dilute Fermi gas (for comparison,  a nucleon
moving through a
nucleus with wave vector
$\bf k$ has an  energy
$\approx V_H + T_k$, which has a gap at $k = 0$). For a uniform Bose gas,
the transition between the phonon and
free-particle spectrum occurs at a wavenumber
\begin{equation}k_0 = \sqrt{8\pi an_0 }=\sqrt2\,ms/\hbar\label{kzro}\end{equation}
obtained by setting $T_{k_0} =V_H$.   The interactions  must be
repulsive ($a>0$)  to ensure that $s$ and $k_0$ are real; in addition, the
interactions  play a
crucial physical role, for $k_0$ vanishes if $a \to 0$.  As noted by
Bogoliubov (based on Landau's
quasiparticle picture), the critical velocity $v_c$ for the destruction of
superfluidity is
here simply the speed of sound,
$s\propto \sqrt{an_0}$, again emphasizing the need for  interparticle
repulsion (in
particular,
$v_c=0$ for a uniform ideal Bose gas).

In addition to the energy eigenvalue, Bogoliubov's canonical transformation
determines the
corresponding eigenstate, which involves a Bose ``quasiparticle'' created
by the operator
\begin{equation}\gamma^\dagger_{\bf k} \equiv u_k\,c_{\bf k}^\dagger +
v_k\,c_{-{\bf
k}}.\label{qp}\end{equation}
  This description looks very
similar to the familiar (and later)
BCS theory of superconductivity \cite{Tink}, but here the
``coherence factors'' satisfy the condition
\begin{equation}u_k^2 - v_k^2 = 1\end{equation}
to ensure that the quasiparticle operators obey Bose commutation relations.
In the interacting ground
state, Bogoliubov showed that  the occupation number $N_k'$ of the state
with wave vector $\bf k$ is
simply
$v_k^2=\frac{1}{2}[E_k^{-1}(V_H + T_k) -1]$, with the limiting forms
\begin{equation}N_k' \approx \cases{k_0/\sqrt8\,k \gg 1,& for $k\ll k_0$, \cr
					                        k_0^4/4\,k^4\ll 1,&for $k\gg k_0$.\cr}
\end{equation}
The total fractional ground-state depletion of the condensate follows by
summing over all nonzero plane
waves:
\begin{equation}\frac{n'}{n} = \frac{1}{n}\int \frac{d^3k}{8\pi^3}\,v_k^2 =
\frac{8}{3}\,\sqrt{\frac{n_0a^3}{\pi}}.\label{depl}\end{equation}
 It is important to recall that this {\it finite\/}
zero-temperature depletion arises  solely  from the presence of the
repulsive interactions.  At
finite temperature,  thermal fluctuations induce additional depletion of
the condensate \cite{FW3},
analogous to that for an ideal Bose gas.
Equation (\ref{depl}) shows that the Bogoliubov  condition  of small
depletion ($N'\ll N$)  thus
requires
$\sqrt{n_0a^3}
\ll 1$;  this depletion is nonanalytic in the scattering length $a$ and
cannot be
obtained with any finite order in perturbation theory (this is the
principal advantage of the canonical
transformation).  A combination with Eq.~(\ref{kzro}) yields the
alternative  condition
$k_0a\ll1$ for a dilute Bose condensate.

	An equivalent  hydrodynamic  description characterizes the same basic
physics. In this case, the operator
\begin{equation}\rho_{\bf k}^\dagger \approx \sqrt{n_0}\,(c_{\bf k}^\dagger +
c_{-\bf k})\end{equation}
creates a density fluctuation with wave vector $\bf k$; it has the same
energy
$E_k$ given in Eq.~(\ref{en}), even though it is a different linear
combination than the quasiparticle
operator in Eq.~(\ref{qp}).  It is easy to show that these ``phonons'' are
longitudinal and hence
irrotational; as discussed below,  vorticity occurs only for condensates
containing discrete vortex
lines.

\section{NONUNIFORM CONDENSATE}
Gross and Pitaevskii \cite{EPG,LPP,FW3}  extended Bogoliubov's theory
to describe a rectilinear vortex line in an otherwise   uniform dilute Bose
gas.  The condensate
density $n_0({\bf r})$ is now spatially varying, and it convenient to
introduce a ``condensate wave
function'' $\Psi({\bf r})$ whose squared absolute value is just the
condensate density $|\Psi({\bf
r})|^2 =  n_0({\bf r})$. Evidently, the total number of condensed particles
is given by $N_0 = \int
d^3r\,|\Psi({\bf r})|^2$, thus fixing the normalization.  In the present
context of trapped alkali
atoms, the Gross-Pitaevskii method must be generalized to include the trap
potential $V({\bf
r})$;  although the actual traps are anisotropic, the present treatment
will consider only the
simpler case of an isotropic harmonic trap with
\begin{equation}V({\bf r}) = {\textstyle
\frac{1}{2}}m\,\omega_0^2\,r^2.\end{equation}

	In contrast to the situation for a uniform condensate, there are now {\it three\/} separate
contributions to  the total energy:  the kinetic energy
\begin{equation}T = -\frac{\hbar^2\nabla^2}{2m},\end{equation}
the  spatially varying Hartree energy of a single particle with the
nonuniform condensate
\begin{equation}V_H({\bf r}) = n_0({\bf r})V_0 = \frac{4\pi a
\hbar^2}{m}\,n_0({\bf r}),\end{equation}
along with the trap potential $V({\bf r})$.  The condensate wave function
satisfies a nonlinear
Schr\"odinger (Gross-Pitaevskii) equation
\begin{equation}(T + V + V_H)\Psi = \mu \Psi,\label{GP}\end{equation}
where $\mu$ is the chemical potential (given at $T = 0$~K by $\mu =
\partial E/\partial N$). The
nonlinearity arises from the Hartree potential, proportional to $|\Psi|^2$.
Equation (\ref{GP}) is
formally like the ``Ginzburg-Pitaevskii'' equation \cite{Ginz}, but its
interpretation is very
different.  Here, the Gross-Pitaevskii equation describes a nonuniform Bose
condensate with small
depletion at low temperature, whereas the Ginzburg-Pitaevskii equation is
analogous to the
Ginzburg-Landau theory of the second-order normal-superconducting phase
transition \cite{GL} and thus
holds only near the transition temperature, where the (thermal) depletion
is necessarily large.

	The trap potential introduces an additional length scale $d_0$ associated
with the zero-point
motion in the trap;  setting $\hbar\omega_0 = \hbar^2/md_0^2$ yields  the
familiar oscillator length
\begin{equation}d_0=\sqrt{\hbar/m\omega_0}.\end{equation}
For comparison, it is convenient to use  the scattering length $a$ and  the
``coherence'' or ``healing'' length
\begin{equation}\xi\equiv 1/k_0 = 1/\sqrt{8\pi an_0}=
\hbar/\sqrt2\,ms\end{equation}
that characterizes the distance over which the condensate wave function
heals back to its asymptotic
value when subjected to a local perturbation ($\xi$ is analogous to the BCS or
Ginzburg-Landau coherence length in a superconductor \cite{GL}).

In the case of trapped
$^{87}$Rb \cite{And}, these parameters are $a \approx 10~\rm nm$ and
$d_0
\approx {\rm a~few~\mu m}$, so that $a\ll d_0$;  for the typical particle
density $n\approx 10^{20}~\rm
m^{-3}$, the coherence length is
$\xi\approx {\rm a~few\times 0.1~\mu m}$.  If
$\xi
\gg d_0$, then the healing length is large compared to the trap dimension
and   the
system is nearly ideal; in this case, the selfconsistent condensate wave
function is the Gaussian
ground state $\Psi_G$ of the harmonic oscillator (the corresponding
condensate density $n_0=|\Psi_G|^2$
is also Gaussian).

 For  most experiments, however, the opposite condition
$a\ll\xi\ll d_0$ applies, and the system is dilute and interacting rather
than ideal, so that the
interparticle repulsion expands the condensate significantly;  in this
limit,  the kinetic energy
operator has only a  small effect on Eq.~(\ref{GP}), leading to an
algebraic equation
\begin{equation}(V + V_H - \mu)\Psi_{TF} = 0.\end{equation}
The resulting  ``Thomas-Fermi'' (TF) approximation for the condensate
density \cite{GSL,HS,BP} yields
\begin{equation}\frac{4\pi a \hbar^2}{m}\,n_{0TF}({\bf r}) = \mu - V({\bf
r})\end{equation}
wherever the right-hand side is positive and zero elsewhere. In the present
case of an isotropic
harmonic potential,   the TF condensate density has an inverted parabolic
profile $n_{0TF}(r) \propto
(R^2-r^2)\,\theta(R-r)$, where
$\theta$ denotes the unit positive step function,  $R/d_0\approx
(15Na/d_0)^{1/5}\gg 1 $ characterizes
the ``condensate radius'' $R$,  and $\mu \approx  \frac{1}{2} \hbar\omega_0
R^2/d_0^2$ is the chemical
potential
\cite{BP}. It is
often preferable to define  the dimensionless ratio
$Na/d_0$, which  is large (small) for dilute interacting (ideal) Bose gas.
 As an aside,
 bulk liquid $^4$He is ``dense,'' for it has  $\xi \approx a$.

\section{SMALL-AMPLITUDE EXCITATIONS OF A NONUNIFORM CONDENSATE}

To treat the small-amplitude excited states in a trap, it is necessary to
generalize
the Bogoliubov quasiparticles to the case of a nonuniform condensate (this
approach was originally
developed for a vortex in a dilute Bose gas \cite{LPP}, but the inclusion of  the trap
potential is straightforward).  Formally, introduce an operator
$\phi^\dagger({\bf r})$ that
creates a noncondensate particle at $\bf r$ [it is simply the fundamental
second-quantized field
operator $\psi^\dagger({\bf r})$ with the condensate operator and wave
function $\Psi({\bf r})$
removed].   This particle operator  can be expressed as a linear
combination of  quasiparticle operators
$\gamma^\dagger_j$ and $\gamma_{-j}$ [compare
Eq.~(\ref{qp})], where the index $j$ denotes one of a set of complete
normal modes and $-j$ denotes
the time-reversed set of quantum numbers \cite{FW4}.  The corresponding
expansion coefficients are  the
Bogoliubov amplitudes (analogous to  wave functions)
$u_j({\bf r})$ and
$v_j({\bf r})$ that obey the coupled linear
 Bogoliubov eigenvalue equations \cite{LPP,AP}
\begin{eqnarray}\big(T + V + 2V_H-\mu\big)u_j - V_Hv_j = E_ju_j,\\
     - V_Hu_j + \big(T + V + 2V_H-\mu\big)v_j  = -E_jv_j,\end{eqnarray}
where the condensate density plays the role of a nonuniform potential
through the Hartree contribution
$V_H$. (The similar coupled equations for a superconductor  are
usually known as the Bogoliubov-de Gennes equations \cite{dG,BdG}.)  The
associated eigenvalues $E_j$
determine the quasiparticle energies  in the presence of the nonuniform
condensate, analogous to the
Bogoliubov energies in Eq.~(\ref{en}) for plane waves.

An equivalent approach is to introduce  hermitian small-amplitude
hydrodynamic operators
\cite{ALF,WG}
\begin{eqnarray}\rho({\bf r}) = \sqrt{n_0({\bf r})} \,[\phi({\bf r}) +
\phi^\dagger({\bf r})],\\
      \Phi({\bf r}) = \frac{\hbar}{2mi\sqrt{n_0({\bf r})}}\,[\phi({\bf r}) -
\phi^\dagger({\bf r})]\end{eqnarray}
 for a density fluctuation and a velocity potential fluctuation at
the point $\bf r$ [the associated velocity is simply ${\bf v(r)} = \nabla
\Phi({\bf r})$].  These
hydrodynamic operators are merely linear combinations of the noncondensate
field operators with known
nonuniform coefficients, and the corresponding normal-mode amplitudes are
similarly linear combinations
of the Bogoliubov amplitudes.  In the present case of a stationary
condensate (the generalization to a
condensate with a persistent current is straightforward \cite{ALF}), the
hydrodynamic  operators obey
the coupled linear equations
\begin{equation}\frac{\partial \rho}{\partial t} + \nabla \cdot \big(
n_0\nabla \Phi\big) =
0,\end{equation}
     \begin{equation}\frac{\partial \Phi}{\partial t} + \frac{4\pi a
\hbar^2}{m^2}\rho +
\frac{\hbar^2}{4m^2n_0}\bigg[\nabla\cdot \bigg(\rho\,\frac{\nabla
n_0}{n_0}\bigg) - \nabla^2\rho
\bigg]= 0\label{Bern}.\end{equation}
The first is the familiar linearized equation of particle conservation, and
the second is a
linearization of the classical Bernoulli's theorem for a compressible
irrotational isentropic fluid
obtained from the appropriate quantum energy-density functional \cite{ALF}.
The corresponding
hydrodynamic amplitudes obey similar coupled eigenvalue equations that
determine the normal-mode
frequencies $\omega_j = E_j/\hbar$.

The standard theory of linear response \cite{FW5} shows that the
density-density correlation function
characterizes  the response of the condensate to a weak external
perturbation that
couples to the density (experimentally, this coupling can be obtained for a
trapped Bose gas by
modulating the curvature of the external trap \cite {Jin,MIT}).  Since the
density operator is here
merely a linear combination of the original particle operators (this
special form reflects the presence
of a macroscopic Bose condensate, and the situation in a Fermi gas is very
different), the resonant
frequencies of the density-density correlation function are necessarily the
same as those of the
Bogoliubov equations, which provides an experimental method to determine
the quasiparticle eigenvalues
$E_j$.

If treated exactly,  the Bogoliubov and the hydrodynamic  descriptions both
yield the same physical
information about the eigenvalues and eigenfunctions.  The former relies on
a quantum formalism with
simple boundary conditions as $r \to\infty$, and the condensate density
appears solely as a
multiplicative coefficient.  In contrast, the latter involves various
spatial derivatives of
the condensate density, but it offers the advantage of physical  intuition
(at least from a classical
perspective), for  the hydrodynamic amplitudes  are like the normal modes
of a selfgravitating
nonuniform star.  Indeed, the recent  field of  ``helioseismology'' uses
the  observed normal modes of
the sun to infer the (otherwise invisible) behavior below the solar
surface; similar studies may be
applicable to the trapped alkali Bose condensates.

\section{RECENT THEORETICAL STUDIES}

The last year has seen the appearance of many theoretical papers and
preprints on various aspects Bose
condensation in dilute trapped gases \cite{BEC}; the present section can
only summarize a few selected
studies that illustrate the recent trends.

\subsection{Stationary condensate}

The GP equation  (\ref{GP}) for a  spherical static Bose condensate in an
isotropic harmonic trap
reduces to a nonlinear ordinary differential equation that can be solved in
many ways, and the
generalization to anisotropic traps is not difficult. A collaborative group from
Georgia~Southern/NIST/Oxford
\cite{EB,RHBE,EDCRB} has performed extensive numerical studies of the
radial condensate wave function
(see also \cite{DS}), including the dependence on the dimensionless parameter
$Na/d_0$ for both positive and negative values (the latter corresponds to
an attractive interaction).
In addition, analytical studies have treated the same problem variationally
\cite{IM,Var}, using trial
functions that interpolate smoothly between the (ideal) Gaussian limit (for
$Na/d_0 \ll 1$) and the
dilute (TF) limit ($Na/d_0\gg 1$).

Usually, the most relevant physical limit is $Na/d_0 \gg 1$, when the TF
approximation describes the
spatial variation of the  condensate density except in a thin layer near
the surface, where the
condensate kinetic energy becomes significant.  Dalfovo, Pitaevskii, and
Stringari have recently
studied this problem analytically using  boundary-layer techniques \cite{DPS}.

\subsection{Rotating condensate (vortices)}
The existence and behavior of vortices in rotating superfluid $^4$He has
long been studied
(see, for example, \cite{BK,Don}).  In particular, the superfluid velocity
${\bf v}_s({\bf r})$ is
proportional to the gradient of the phase of the condensate wave function.
As a result,  the superflow
is irrotational ``almost'' everywhere except in discrete singular regions
of dimension  $\sim \xi$ (the
nodes of
$\Psi$) where the phase is undefined;  superfluid  vorticity (namely, the
regions where
${\bf \nabla}\times {\bf v}_s \neq 0$) is localized in  the cores of the
superfluid vortices. In
addition,
$\Psi$ is  single-valued  whenever its coordinate $\bf r$ traverses a
closed path in the fluid, which
ensures
 the quantization of circulation $\kappa$  in units of
$h/m$.

The simplest system for liquid $^4$He is a long circular cylindrical
container with radius $R$ that can
rotate about its symmetry axis. For small angular velocity $\Omega$, the
fluid remains at rest with no
vortices, but, at a critical value $\Omega_{c1} \sim
(\hbar/mR^2)\,\ln(R/\xi)$, it becomes favorable to
create one singly quantized vortex line on the axis of the cylinder.
Similar questions have been
considered for axisymmetric harmonic traps in the context of   Bose
condensed alkali atoms
\cite{BP,EDCRB,DS}, where, for current experiments on $^{87}$Rb, the
critical angular velocity should
be
$\Omega_{c1}\sim 50 \rm\ rad/s$.

\subsection{Excited states}
As noted in the paragraph below Eq.~(\ref{Bern}), the linearized normal
modes of the condensate
characterize both its small-amplitude dynamics and its linear response to a
weak external perturbation
that couples to the particle density.  As a result, considerable
theoretical effort has been devoted to
the calculation of these normal modes, both numerically \cite{ERBDC} and
analytically  in the TF
limit of a large condensate \cite{S}.  For an  axisymmetric trap, the eigenstates can be labeled with
the azimuthal quantum number $m$. In  recent  experiments \cite{Jin,MIT},
the trap
curvature was modulated at a prescribed external frequency, and the
observed resonant behavior
measures the appropriate eigenfrequencies.   Detailed comparison between
theory and experiment yields
good agreement  for the lowest states with $m = 0$ and $m = 2$.

\section{SOME OPEN QUESTIONS}

This field is developing rapidly, so that any selection of open questions
is somewhat arbitrary (and
sure soon to be   irrelevant).  Nevertheless, the following few topics are
likely to remain
of interest for at least the immediate future.

\subsection{Normal modes of a large condensate}
In the limit of a large condensate ($Na/d_0 \gg 1$), the TF approximation
for the condensate
density holds apart from the surface region.  An incompressible spherical
fluid has surface waves as
its  lowest normal modes, but it is less clear what happens for a dilute
nonuniform Bose gas.  In
principle, numerical analysis can yield all the  normal modes for a
particular configuration,
but analytical results  would be helpful to gain some insight into the
relevance  of  various
physical parameters and the interplay between bulk modes  and surface modes.

\subsection{Axisymmetric rotating condensate}

From an experimental viewpoint,  the  simplest  way to
create   vortex lines in liquid $^4$He is to cool the sample while it
rotates, so that the onset of
the
 superfluid state occurs at fixed $\Omega$. This method works even for a
circular cylindrical
container, because the rotating walls entrain the viscous normal fluid
above $T_\lambda$,
thus creating the  superfluid in a state of rotation.

An alternative procedure is to cool a stationary container of $^4$He  below
$T_\lambda$ and then spin it
up to some final angular velocity $\Omega$.  For $\Omega {\
\lower-1.2pt\vbox{\hbox{\rlap{$<$}\lower5pt\vbox
{\hbox{$\sim$}}}}\ }\Omega_{c1}$, the superfluid typically
remains at rest, but for $\Omega {\
\lower-1.2pt\vbox{\hbox{\rlap{$>$}\lower5pt\vbox
{\hbox{$\sim$}}}}\ }\Omega_{c1}$, the inevitable surface roughness of the
rotating wall
apparently  initiates the formation of a vortex.  Unfortunately, the
detailed mechanism for vortex creation remains unclear, and the observed
hysteresis implies
  significant metastability \cite{PS}.

In liquid $^4$He, only singly quantized vortices are expected
theoretically, in agreement with many
different  experimental observations \cite{Don,PS}.  Imaging the
vortices in rotating superfluid $^4$He  remained elusive for many years,
in contrast to the
corresponding success with
 quantized flux lines in type-II superconductors \cite{GL}.  Nevertheless,
clever use  of  ion
trapping on the vortex cores \cite{BK,Don} eventually yielded remarkably
explicit images of the vortex
positions \cite{YGP,YP}, which are really just the nodes in the condensate
wave function.  Similar
explicit images of vortices in dilute trapped alkali Bose condensates would
be very desirable (it is not
even obvious that only singly quantized vortex lines will occur).

\subsection{Nonaxisymmetric rotating condensate}
As noted above, spinning up    a  long  circular cylinder of
superfluid
$^4$He from rest frequently involves hysteresis, and it is not clear how a
similar procedure for
    condensed alkali atoms in an axisymmetric trap can succeed in creating
vortices (in the
absence of physical boundaries, what does it mean to rotate the trap?).
One obvious possibility is to
use  an {\it asymmetric\/} container (for example,  different oscillator lengths
$d_x$ and $d_y$ in the $xy$ plane, with the harmonic trap rotating about
the $z$ axis).

 A similar
question has been treated theoretically for an incompressible superfluid in
a long cylinder with either
elliptical or rectangular  profile \cite{Def}.  The principal new feature
is that the rotating walls
impel the superfluid into irrotational motion {\it even in the absence of
vortices\/}.  Indeed, for an
elliptical profile with semiaxes $a$ and $b$, the resulting  induced  angular momentum is reduced
relative to that for solid-body rotation by a factor
$(a^2-b^2)^2/(a^2+b^2)^2$ \cite{Def}, which
 vanishes for a circular profile but is otherwise positive and less than
one. In addition, the
critical angular velocity
$\Omega_{c1}$ for the onset of vortex formation is increased relative to
that for a circular
cylinder for  the following physical reason.  It is easy to show that solid-body
rotation $\bf \Omega\times r$  is the true  equilibrium of an unconstrained
rotating incompressible
fluid in a container, but such a configuration is forbidden to a superfluid
because ${\bf v}_s$ must
remain irrotational almost everywhere (recall that  solid-body rotation has
uniform nonzero  vorticity
$2\bf\Omega$).  A superfluid in a rotating circular cylinder has no purely
irrotational motion and can
mimic  solid-body rotation only by creating  singly quantized vortices;  in
contrast, a  rotating
elliptic cylinder does have  irrotational flow that already acts somewhat
like  solid-body rotation,
reducing the need to create one or more discrete vortices.  Careful
experiments confirm this prediction
in considerable detail \cite{DCP}.

\subsection{Excited states of a vortex}
A classical rectilinear vortex with circulation $\kappa$ in an
incompressible fluid has
oscillatory normal modes that  propagate along (and  are confined to) the
axis, with a long-wavelength
dispersion relation
$\omega\approx (\kappa k^2/4\pi)\,\ln(1/k\xi)$, where $k$ is the axial
wavenumber and $\xi$ is the core
size (assuming $k\xi \ll 1$) \cite{BK,Don}.  Similar vortex waves occur for
a vortex in an unbounded
dilute Bose condensate \cite{LPP,Vort}, where $\kappa = h/m$. In addition,
the phonon-like modes of a
uniform Bose gas are modified by the presence of a vortex, leading to
asymptotic phase shifts. The
resulting phonon-vortex  scattering contributes to the mutual friction, as
observed
in rotating superfluid
$^4$He through an  excess attenuation of second sound relative to that for
stationary
superfluid \cite{BK,Don}.

The situation for one or more vortices in a
trapped dilute Bose condensate will necessarily  differ considerably from
that in an unbounded medium.
As one simple example,   the ratio of the condensate radius $R$  to the
vortex core
$\xi$ is typically $\sim 10$, so that only  relatively few vortices can
form before the cores
overlap and the system becomes ``normal'' (a finite-size analog of  the
destruction of
superconductivity at the upper critical field
$H_{c2}$ \cite{GL}).  A fundamental question concerns the various possible
normal modes that can
occur for vortex lines in these relatively small dilute trapped systems.
Although there will
certainly be   phonon-like modes  that are scattered by the vortices, it is
less obvious that such a
system can support an analog of the bound vortex-wave modes.

%
% Place here the list of the references:
%

\end{document}